\documentclass{article}
\usepackage[a4paper, total={7in, 10in}]{geometry}

\usepackage[utf8]{inputenc}
\usepackage{booktabs}
\usepackage{graphicx}
\usepackage{rotating}
\usepackage[hyphens]{url}
\usepackage{todonotes}
\usepackage{tikz}
\usepackage{caption}
\usepackage{subcaption}
\usepackage{float}
\usepackage{xcolor}
\definecolor{lblue}{HTML}{D8E8FB}
\definecolor{dblue}{HTML}{668EBC}
\definecolor{lviolet}{HTML}{e2d5e6}
\definecolor{dviolet}{HTML}{9973a4}
\definecolor{lgreen}{HTML}{d3e8d5}
\definecolor{dgreen}{HTML}{7cb36b}
\definecolor{lyellow}{HTML}{fff2cf}
\definecolor{dyellow}{HTML}{e2c276}
\definecolor{lorange}{HTML}{ffe6ce}
\definecolor{dorange}{HTML}{edbd66}

\providecommand{\keywords}[1]{\textbf{\textit{Index terms---}} #1}
\begin{document}

\title{Who Watches the New Watchmen? The Challenges for Drone Digital Forensics Investigations}

\author{Evangelos Mantas$^{1}$ and Constantinos Patsakis$^{1,2}$\\
        \small $^{1}$Department of Informatics, University of Piraeus,\\\small 80 Karaoli \& Dimitriou str., 18534 Piraeus, Greece\\
        \small $^{2}$Information Management Systems Institute,\\\small Athena Research Center, Artemidos 6, Marousi 15125, Greece\\
}
\date{}




\maketitle
\begin{abstract}
 The technological advance of drone technology has augmented the existing capabilities of flying vehicles rendering them a valuable asset of the modern society. As more drones are expected to occupy the airspace in the near future, security-related incidents, either malicious acts or accidents, will increase as well. The forensics analysis of a security incident is essential, as drones are flying above populated areas and have also been weaponised from radical forces and perpetrators. Thus, it is an imperative need to establish a Drone Digital Forensics Investigation Framework and standardise the processes of collecting and processing such evidence.

 Although there are numerous drone platforms in the market, the same principles apply to all of them; just like mobile phones. Nevertheless, due to the nature of drones, standardised forensics procedures to date do not manage to address the required processes and challenges that such investigations pose. Acknowledging this need, we detail the unique characteristics of drones and the gaps in existing methodologies and standards, showcasing that there are fundamental issues in terms of their forensics analysis from various perspectives, ranging from operational and procedural ones, and escalate to manufacturers, as well as legal restrictions. The above creates a very complex environment where coordinated actions must be made among the key stakeholders. Therefore, this work paves the way to address these challenges by identifying the main issues, their origins, and the needs in the field by performing a thorough review of the literature and a gap analysis.
\end{abstract}


\section{Introduction}
A few years ago, drones were exclusively used by government and military organisations, but the ever-evolving technology made them accessible to the public. What was labelled as a ``technology of the future'', quickly become a technology of today, as drones integrated to the needs of the market rapidly. Drones evolved from toy machines used by numerous flight enthusiasts to machines that can undertake operations on everyday tasks. From seemingly simple applications like high-level quality infrastructure inspections, and everyday goods deliveries to life-saving Search and Rescue (SAR) missions, drones have established their value to the society. The revolutionary use of drones comes with side effects as malicious actors have also evolved to incorporate drones in their \textit{modus operandi} as they are cheap, reliable and easy to modify to carry payloads (e.g. drugs, explosive devices). Over the last five years, drones have been used to disrupt airport operations\footnote{\url{https://www.forbes.com/sites/grantmartin/2019/01/08/londons-heathrow-airport-briefly-closed-following-possible-drone-sighting/}}, attempted high-value persons assassinations\footnote{\url{https://www.bbc.com/news/world-latin-america-45073385}}, and used in attacks against industry complexes\footnote{\url{https://www.nytimes.com/2019/09/14/world/middleeast/saudi-arabia-refineries-drone-attack.html}} to name a few. The number of malicious acts is expected to rise since perpetrators are becoming more acquainted with this new technology.

The need to address this emerging technology has pushed Digital Forensics Science to further specialise in handling incidents involving drones and their supporting equipment or devices. Since this need is relatively new, there is a limited amount of information and specialisation on the matter, and as there is no standardised practice to conduct a forensic investigation around drones, the investigator relies on already established procedures, relevant to other digital devices, that may not be applicable to a drone. Furthermore, the presence of a large number of different devices, that support the operation of a drone expands the complexity of the investigation, as the forensics expert will have to handle the supporting devices as well as to find and correlate evidence regarding the drone operation.

This work focuses on explaining the reason behind the need for a standardised procedure by identifying the related devices needed to be examined and the evidence stored in them. Therefore, we review the related literature, highlighting key differences with traditional digital forensics operations and the challenges that these introduce. We argue that since both drone accidents and malicious acts will become increasingly frequent in the near future, a forensics framework to handle such incidents should be set in a timely manner. In this regard, our work performs a thorough literature review which covers the work in both academic and `gray' literature as it spans to, e.g. international standards. Based on this, our gap analysis illustrates the needs in the field and serves as a guideline for both academic and practitioners in the field.

The rest of this manuscript is organised as follows. First, we present the related work by first providing an overview of the drone and the most relevant forensics tools for drones. Then, in Section \ref{sec:drone_evidence}, we discuss which are the primary sources for evidence in a drone and their types, to proceed with a generic overview of the cases that a forensics investigator would have to face. To facilitate her task, we accompany the latter with a comprehensive and prioritised list of evidence for each case. In Section \ref{sec:gaps}, we first present the relevant forensic models and standards to discuss the primary challenges in the forensics investigation in the form of a gap analysis afterwards. The key differences of drones and their particularities indicate that we are currently missing not only the needed tools but also the procedures to address the upcoming forensics cases. The article then concludes and streamlines some key findings and open issues for the further development of the field.

\section{Related work}
\label{sec:related}
The goal of this section is to acquaint the reader with the drone and the existing tools to collect forensic evidence from a drone. Therefore, we briefly provide an overview of a drone and proceed to present some relevant tools and their scope.

\subsection{Overview of a drone}
Unmanned Aerial Vehicles are commonly acknowledged as drones in everyday life. It is important to make a distinction between the systems with autonomous capabilities and the hobbyist flying machines; the capabilities of which rely only on the pilot's skills. Since these radio-controlled (RC) machines require minimal and basic hardware to fly, usually a `flight board` to receive and process the signal to spin the motors, they do not offer any tangible digital forensics. Therefore, this work focuses on drones with autonomous capabilities. The primary components of a drone are the following:
\begin{itemize}
\item \textbf{Flight Controller:} The most critical component of a drone since it contains a variety of sensors (e.g. accelerometer, gyroscope) and software, or as widely know as the autopilot, to allow the execution of autonomous commands and more complex flight scenarios. The flight controller can also receive input from sensors (e.g. lidar, infrared) and adjust the flight path to avoid collision with nearby objects without any human interaction.

\item \textbf{Companion Computer:} The flight controller can be directly connected to a small on-board companion computer (e.g. Raspberry Pi, Asus Tinker-board) that augments the drones' capabilities even further, running a real-time Operating System (OS). A task like an object avoidance can be executed by combining the on-board range-finding sensors connected on the companion computer and provide that feedback to the flight controller. Companion computers are expected to be found on makeshift/DIY flying machines or very sophisticated commercial drones to handle demanding tasks, such as object avoidance and 3D object reconstruction.

\item \textbf{Electronic and Hardware Components:} Electronic speed controllers (ESC), power distribution board, battery, propellers and motors, are the essential components to provide power, lift and propulsion.

\item \textbf{GPS Receiver:} Used to provide the location of the drone and enables features such as Return To Home/Land (RTH/RTL) and autonomous mission's flight path.

\item \textbf{Radio Frequency Receiver:} Used to receive control signals from the ground-based transmitter, either the pilot or a ground control station (GCS).

\item \textbf{Body:} The main fuselage of the UAV used to protect and house all the components.

\item \textbf{Payload-Sensors:} As the industry becomes more relied on drones, cameras and sensors (e.g. lidar, range-find sensor) are deployed on drones becoming an essential part of the system since these are the ones offering the main functionality.
\end{itemize}

\subsection{Drone forensics tools}
The need to forensically investigate drones has sparked great interest among cybersecurity researchers. Early work focused on identifying the crucial segments related to a drone investigation \cite{kovar_dominguez_murphy_2016}. In their work Kao et al. \cite{KAO20191890} and Yousef et al. \cite{9035365}, \cite{9079012} have used specific models of drone maker DJI as case studies to analyse evidence recorded in drone flights. The same approach, to use a particular drone model from another popular manufacturer, Parrot Bebop, was used in the work of Kamoun et al. \cite{parrot-forensics}. In their research, Clark et al. \cite{DROP} used the DJI Phantom III as their case study to create an open-source tool (DROP) to extract data from the drone, that is stored in an encrypted, encoded, and proprietary file. Since the release of the research, DROP has been used as the default tool for the analysis of DJI log files, as there is currently no other tool available in the market. In our previous work \cite{Gryphon}, we focused on describing techniques to extract data from drones with Ardupilot flight open-source firmware, developing an open-source tool that is agnostic to the type of the vehicle model or type, extracting data from the MAVLink messages transmitted on the drone. It must be highlighted that most of the drone forensics researches focus on flying vehicles that are popular on the market, running on proprietary software, and although they provide a solution to conduct a forensics investigation, their methodology, or parts of the methodology, cannot be standardised. The techniques described in the research may soon become obsolete or irrelevant due to the changes drone manufactures will most likely make in their future products.

\section{The drone and evidence sources}
\label{sec:drone_evidence}
In this section, we highlight the primary components of a drone that can be used as sources for extracting evidence. Then, we proceed to pinpoint the cases in which a forensics investigation will be required and prioritise, which are the most relevant evidence that an investigator would have to look for per case.

\subsection{Drone from the forensic expert's perspective}
\begin{figure*}[!ht]
 \includegraphics[width=.75\textwidth]{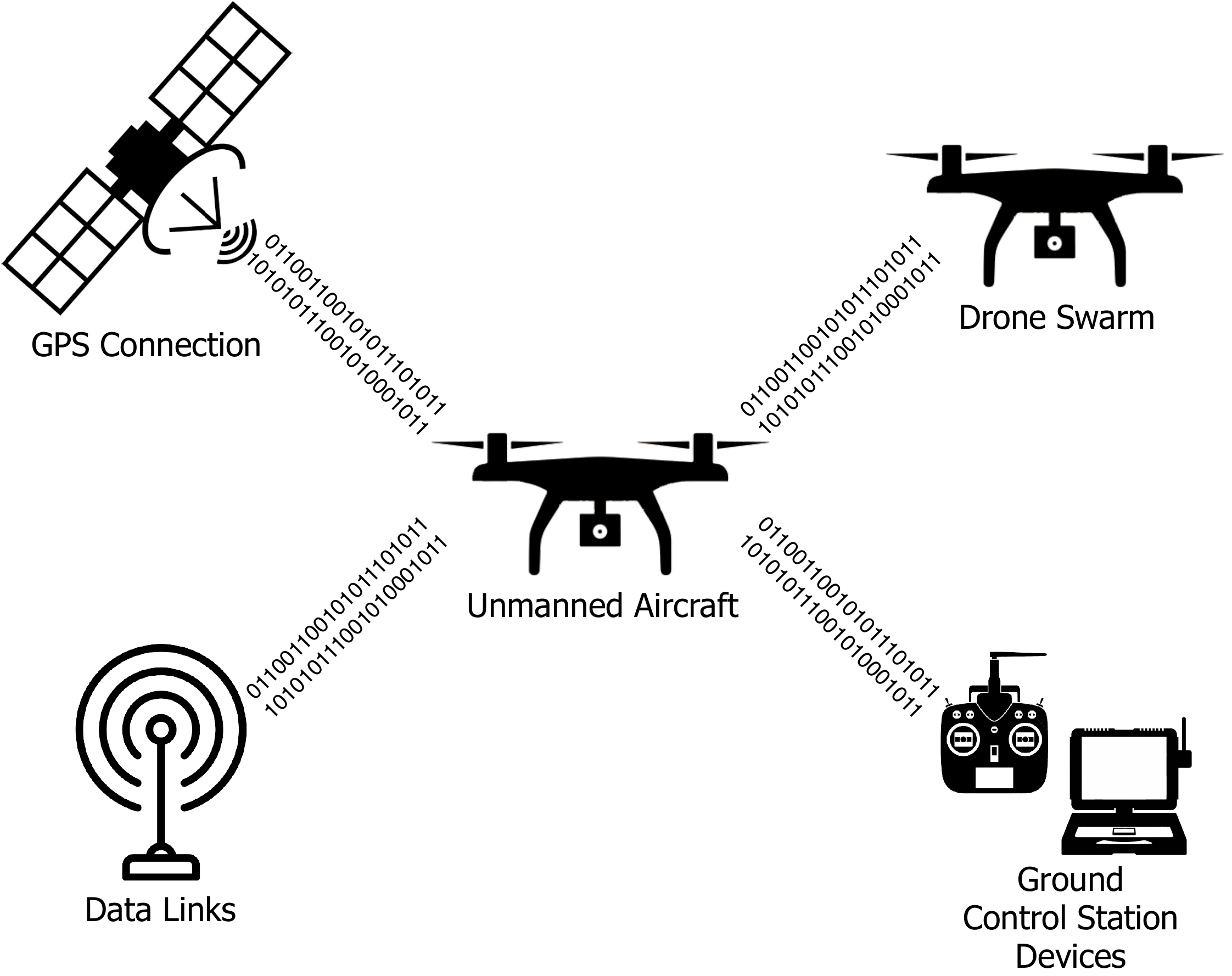}
 \centering
 \caption{Components of an Unmanned Aerial System (UAS).}
 \label{fig:UAS}
\end{figure*}

The growth of drone technology has introduced the use of UAVs to a number of applications in the private and military sector. Drones that deliver payloads (e.g. medical supplies or everyday goods) or drones conducting surveillance over combat areas, use a variety of sensors and systems to communicate and receive commands from the ground operators. From laptops and mobile devices to directional antennas and 5G towers to name a few, the operation of a drone relies on a \textit{system}, the Unmanned Aerial System (UAS). Therefore, the digital forensics expert must focus on the whole UAS and not only the flying vehicle. The key components of the \textit{UAS} (see Figure \ref{fig:UAS}), which may also come in the form of a swarm, are the following:
\begin{itemize}
\item \textbf{Unmanned Aircraft (UmA):} The flying vehicle as described in the previous section, is probably the most critical component since the whole system exists to assist the drone's operation and its autonomous capabilities. It should also be noted that another or multiple flying vehicles may be part of the same UAS, which is described as a \textit{swarm}.
\item \textbf{Ground Control Station (GCS):} Ground Control Station is the ground-based software and hardware that provides communication and enables the remote control of the drone. GCS software can be installed on systems on fixed locations (e.g. military airbases) or in portable systems such as laptops and mobile phones. The acquisition of the GCS device is of great significance since it contains most of the flight data and can be linked with an individual. There are commercially available drones that come equipped with controllers with proprietary software to control the drone, a hybrid device combining the analogue RF control and the digital interface of a mobile platform, the acquisition of which can also contain considerable evidence.
\item \textbf{Communication Data-Link (CDL):} Data links are mainly used for UAV control, video transmission from the on-board camera and telemetry providing information on the location of the vehicle or other sensor values. Each data link broadcasts on a different radio-frequency (RF) wavelength with 4Ghz and 900Mhz mainly being used for \textit{UAV control}, 3-4-5Ghz for \textit{video transmission} and 400-900Mhz for \textit{telemetry}\footnote{\url{https://www.911security.com/learn/airspace-security/drone-fundamentals/drone-communication-data-link}}. Small UAVs (sUAV) may also use Bluetooth and WiFi communications to execute all the previous operations, but their flight radius is significantly reduced.
\end{itemize}

All these interconnected components store data related to the drone operation or serve them to other devices. These data can be classified in the following types:
\begin{itemize}
\item \textbf{Flight logs and predefined flight missions:} As mentioned before, the \textit{flight controller}, is responsible for the autonomous capabilities, has another function during the flight. It generates a log that contains information about critical sensors like the GPS position and altitude, resulting in a reconstruction of the flight path, providing valuable information about the drone activity to the forensics investigator. Moreover, they contain key equipment data and error logs. For instance, battery and signal statistics which along with other error messages may point to a malfunction. The pilot of the drone also can create a fight mission that can be loaded on the flight controller directly. This file can provide information about previous flights that were not stored or deleted, or even scheduled flights.
\item \textbf{Media:} Drones equipped with cameras offer a vast source of images and video footage captured during the flight. The acquisition of such media can indicate the target of the drone's mission (e.g. spying on the industrial or military complex) and the areas affected by the drone's flight. These multimedia may contain frames of the drone operator, footages before the collision, resulting in a quicker identification by the forensics investigator.
\item \textbf{Sensor or other payload logs:} The payload of a drone is not limited to cameras only. Drones are becoming part of our everyday life and undertake operations that require advanced hardware on board (e.g. 3D reconstruction of the ground, infrastructure inspection). Although this kind of equipment captures large quantities of data, they may not have any significant forensic value to the investigation, as data collected are to be utilised for a specific business operation.
\end{itemize}

\begin{figure*}[!ht]
 \includegraphics[width=\textwidth]{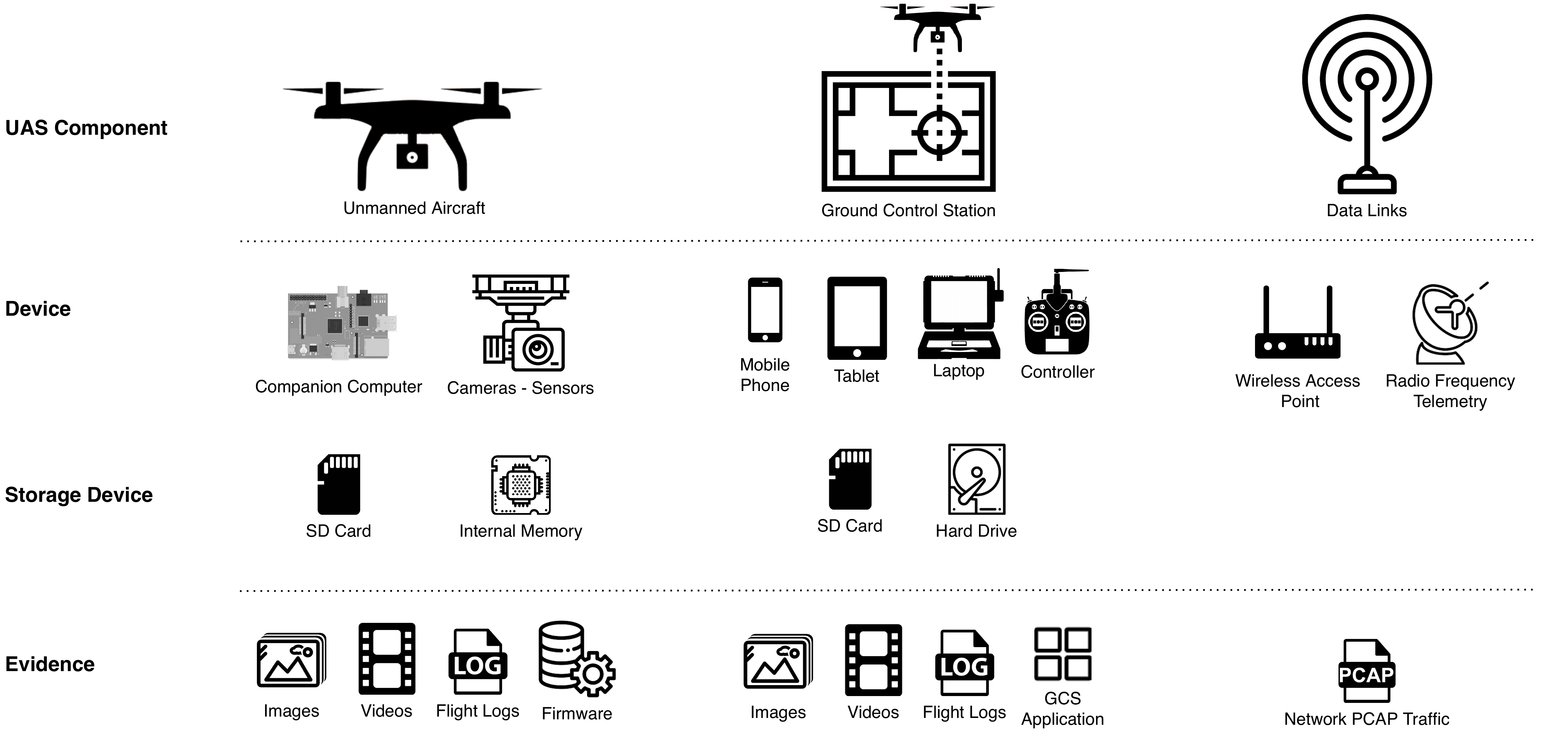}
 \centering
 \caption{Evidence Distribution on UAS devices.}
 \label{fig:UAS-D-E}
\end{figure*}

The presence of a large variety of devices also provides a large source of evidence collection. Each device in the UAS will most likely have a storage unit where the drone forensics investigator can uncover details about the activity of the drone and its operator. The presence of multiple evidence on different devices means that if the drone or a device of the UAS is rendered unrecoverable due to, e.g. a damage or anti-forensics techniques\footnote{\url{https://blog.eccouncil.org/6-anti-forensic-techniques-that-every-cyber-investigator-dreads/}} (e.g. encrypted internal drone memory), there is a high possibility that other evidence can be retrieved. The most common devices that can contain evidence of a drone flight are described below:

\begin{itemize}
\item \textbf{Drone Internal Storage:} The drone's internal on-board storage contains, although in a limited amount due to hardware limitations, as it rarely exceeds tens of GBs, flight logs and media captured during the flight. Usually, the internal memory serves as an alternative or ``backup'' location to store data if removable storage is not present.

\item \textbf{On-board Removable Storage:} The widespread need to capture a large amount of data can be satisfied by using removable media like the SD card. Provided with the fact that storage capacities of micro SD cards have exceeded the TB mark, the low price per byte ratio has established them as the dominant data storage solution in UAVs. Most of the drones have an SD card slot to save images and video footage recorded during the flight or even the ``hard drive'' of the companion computer.

\item \textbf{Companion Computer Storage:} As mentioned before, companion computers are running a real-time OS. This may contain digital evidence as ``raw'' information of drones activity, like the execution of flight-mission scripts and transmission of data to other devices.

\item \textbf{Mobile Devices:} Mobile GCS applications provide the ability to control a drone or receive a payload (e.g. video stream of the drone's on-board camera) via wireless connection (e.g. WiFi, Bluetooth) and should also be considered as a prospective data source that can be directly linked to the pilot.

\item \textbf{Network Packets:} The monitoring and capturing of packets from the UAS network, can lead to the discovery of numerous payloads or devices and servers that the fight data are stored, as an additional source of evidence that could not be otherwise obvious to the forensic investigator.

\item \textbf{Cloud Storage:} There is a currently increasing trend in UAV technology, where manufacturers allow their clients to store their logs and images on their servers. Data storage outsourcing to third parties' servers, internal UAS cloud, and backup servers, due to the demand to reduce local storage media, should also be considered during a UAS forensics investigation.
\end{itemize}
An indicative list of the devices and the evidence found on a typical UAS is illustrated in Figure \ref{fig:UAS-D-E}.

\subsection{Cases for forensics investigation}
It is deemed necessary to outline the cases that a digital forensic investigation would be required, to highlight which evidence might be available and which ones could be more more relevant for each case.

The most ``benign'' cases involve accidents where the forensic investigator should determine its causes. It is clear that even a small malfunction, due to the altitude that a drone may have and the difficulty to balance once something starts falling from the sky, severe damages would be suffered by the drone. Moreover, while one may not care about the damage to the drone and its payload per se, insurance issues due to the caused damages may require the investigation of the accident. It should be noted that weather issues due to sudden and local phenomena or even collisions with a bird or another drone fall into the accident category and have to be investigated. In such cases, the most relevant evidence would be the flight logs and media captured, since they would be available on the pilot's GCS device (e.g. accident in a delivery flight).

Going a step further, we have to consider cases of malicious \textit{physical} acts either from the drone or towards it. On the one hand, we consider physical attacks towards the drone that include but are not limited to, physical damage or theft of the drone or its payload. On the other hand, we consider acts where the drone is used to perform malicious physical acts which include, among others interception, surveillance, theft, delivery of illegal products, and use of armoury.

Finally, we consider digital attacks to and from the drone \cite{yaacoub2020security}. Digital attacks towards the drone can be considered jamming, command injection, electromagnetic pulse (EMP), denial of service (e.g. flooding a network interface), and data leakage through the interception of a vulnerable protocol. Of specific interest can be considered the malware infection of the drone due to the existence of, e.g. a vulnerable interface or service. Moreover, we have to also consider firmware backdoors as more and more supply chain attacks are becoming more relevant \cite{europol2017internet}. On the other side, the drone can be used to launch a digital attack, e.g. infiltrate a wireless network, perform a denial of service of an unprotected service or partially disrupt it.

\begin{table}[th]
 \centering
 \begin{subtable}[h]{\columnwidth}
 \centering
 \begin{tabular}{lc}
 \toprule
 Evidence source& Color \\
 \midrule
 Fuselage* & \tikz\draw[thick,black, fill=gray!50] (0,0) circle (1ex);\\
 GCS Device* & \tikz\draw[thick,black,fill=black] (0,0) circle (1ex);\\
 Flight Logs & \tikz\draw[thick,black,fill=white] (0,0) circle (1ex);\\
 Media & \tikz\draw[thick,dviolet,fill=lviolet] (0,0) circle (1ex);\\
 Sensors & \tikz\draw[thick,dblue,fill=lblue] (0,0) circle (1ex);\\
 On-board Computer & \tikz\draw[thick,dyellow,fill=lyellow] (0,0) circle (1ex);\\
 Payload & \tikz\draw[thick,dgreen,fill=lgreen] (0,0) circle (1ex);\\
 Network Packet Logs& \tikz\draw[thick,dorange,fill=lorange] (0,0) circle (1ex);\\
 \bottomrule
 \end{tabular}
 \caption{Color coding of evidence sources. Note(*): Related physical evidence }
 \end{subtable}

 \begin{subtable}[h]{\columnwidth}
 \centering
 \begin{tabular}{lp{1.5in}}
 \toprule
 Case& Sources priority \\
 \midrule
 Collision/Crash land &
 \tikz\draw[thick,black, fill=gray!50] (0,0) circle (1ex);
 \tikz\draw[thick,black,fill=black] (0,0) circle (1ex);
 \tikz\draw[thick,black,fill=white] (0,0) circle (1ex);
 \tikz\draw[thick,dviolet,fill=lviolet] (0,0) circle (1ex);
 \tikz\draw[thick,dblue,fill=lblue] (0,0) circle (1ex); \\
 Illegal Area Surveillance &
 \tikz\draw[thick,black,fill=white] (0,0) circle (1ex);
 \tikz\draw[thick,dviolet,fill=lviolet] (0,0) circle (1ex);
 \tikz\draw[thick,dblue,fill=lblue] (0,0) circle (1ex);
 \tikz\draw[thick,dgreen,fill=lgreen] (0,0) circle (1ex);
 \tikz\draw[thick,black, fill=gray!50] (0,0) circle (1ex);
 \tikz\draw[thick,black,fill=black] (0,0) circle (1ex);\\
 Ordnance/Payload Delivery &
 \tikz\draw[thick,dgreen,fill=lgreen] (0,0) circle (1ex);
 \tikz\draw[thick,black,fill=white] (0,0) circle (1ex);
 \tikz\draw[thick,dviolet,fill=lviolet] (0,0) circle (1ex);
 \tikz\draw[thick,dblue,fill=lblue] (0,0) circle (1ex);
 \tikz\draw[thick,black, fill=gray!50] (0,0) circle (1ex);
 \tikz\draw[thick,black,fill=black] (0,0) circle (1ex); \\
 Cyber-attack (Victim) &
 \tikz\draw[thick,dyellow,fill=lyellow] (0,0) circle (1ex);
 \tikz\draw[thick,dorange,fill=lorange] (0,0) circle (1ex);
 \tikz\draw[thick,black,fill=white] (0,0) circle (1ex);
 \tikz\draw[thick,dviolet,fill=lviolet] (0,0) circle (1ex);
 \tikz\draw[thick,dblue,fill=lblue] (0,0) circle (1ex);
 \tikz\draw[thick,dgreen,fill=lgreen] (0,0) circle (1ex);
 \tikz\draw[thick,black, fill=gray!50] (0,0) circle (1ex);
 \tikz\draw[thick,black,fill=black] (0,0) circle (1ex);\\
 Cyber-attack (Actor) &
 \tikz\draw[thick,dyellow,fill=lyellow] (0,0) circle (1ex);
 \tikz\draw[thick,dorange,fill=lorange] (0,0) circle (1ex);
 \tikz\draw[thick,black,fill=white] (0,0) circle (1ex);
 \tikz\draw[thick,dviolet,fill=lviolet] (0,0) circle (1ex);
 \tikz\draw[thick,dblue,fill=lblue] (0,0) circle (1ex);
 \tikz\draw[thick,black, fill=gray!50] (0,0) circle (1ex);
 \tikz\draw[thick,black,fill=black] (0,0) circle (1ex);
 \tikz\draw[thick,dgreen,fill=lgreen] (0,0) circle (1ex);
 \\
 Weather related accident &
 \tikz\draw[thick,dblue,fill=lblue] (0,0) circle (1ex);
 \tikz\draw[thick,black,fill=white] (0,0) circle (1ex);
 \tikz\draw[thick,black,fill=black] (0,0) circle (1ex);
 \tikz\draw[thick,black, fill=gray!50] (0,0) circle (1ex);
 \tikz\draw[thick,dviolet,fill=lviolet] (0,0) circle (1ex);
 \\
 \bottomrule
 \end{tabular}
 \caption{Prioritisation of evidence sources according to the case.}
 \end{subtable}
 \caption{High level description of drone forensic investigation cases.}
 \label{tbl:drone_cases}
\end{table}

Using the collected evidence, the investigator has firstly to assess the nature of the incident and whether it is an offence, e.g. whether it is an accident or a malicious act. Based on the scene under investigation, the collected evidence and the visible incurred damages, the investigator has to assess the actual impact as further damages may have been caused. The evidence will allow her to estimate when and where the incident occurred but also to extract some key locations of interest, e.g. from where the drone was launched and where it was heading to. The scheduled flight and the payload may also denote the drone's purpose. For the attribution, the investigator also has to consider the stored data (images, videos), metadata (emails, registered accounts, stored WiFi), hardware tags, as well as wet evidence (fingerprints, DNA) to assess the possible owner and controller of the drone. Clearly, the above highlight the strong connection of the digital evidence with the physical ones which act rather complementary.

To summarise the evidence collection sources, in Table \ref{tbl:drone_cases}, we highlight and prioritise the most relevant evidence sources per case. The table clearly shows how the generic cases we discussed earlier differ as the investigator has to look for different evidence.

\section{Challenges and gaps for drone digital forensics}
There is a constant challenge to stay up to date with the rapid changes in the latest technologies. Drones are beginning to shape tomorrow's society, used by many individuals for both personal and professional purposes. The challenge of carrying out digital forensics on drones and associated devices as part of the UAS, is to identify flight routes, possible malfunctions, collisions, and associated media (pictures and videos) contained within the system's devices. These may facilitate the investigator to understand the mission of the drone, whether the operation was successfully performed, the people or assets that were affected by it, and the reasons behind possible damage or collision. The findings derived from electronic evidence must follow a standard set of guidelines to ensure that the investigation is not challenged in a court of law \cite{mckemmish2008digital}. Moreover, the border-less nature should also be considered due to cloud storage (e.g. evidence on a server hosted in a third country). In fact, as indicated by NIST, cloud forensics are also facing many challenges \cite{herman2020nist} and a draft mapping of, e.g. ISO 27037 is not that simple\footnote{\url{https://downloads.cloudsecurityalliance.org/initiatives/imf/Mapping-the-Forensic-Standard-ISO-IEC-27037-to-Cloud-Computing.pdf}}. Additionally, the provision of the evidence from a cloud provider or a third party, in general, is not necessarily granted, as shown in the case of Apple and the San Bernardino shooting case\footnote{\url{https://edition.cnn.com/2016/02/16/us/san-bernardino-shooter-phone-apple/index.html}}. Note that even if the cloud service provider is willing to cooperate the data may be stored across different servers and countries. The presence of multiple devices and various formats on a UAS requires a high level of expertise from the forensic's experts that take part in the investigation since each device requires different acquisition and examination techniques.

\subsection{Relevant standards, practices and challenges}
\label{sec:gaps}
Standards are a vital necessity to ensure conformance and mutual compliance across the geographical and jurisdictional border. There are currently numerous standards and established practices provided by organisations worldwide using accepted methods. The technical details on how to forensically approach the investigation differs, depending on the device. The analysis of electronic evidence is typically categorised into the following phases: \textit{identification}, \textit{collection}, \textit{acquisition}, \textit{preservation} and \textit{disposal}, although the exact phases naming may vary due to the usage of different forensic model that each organisation finds appropriate for their needs.

Currently, there are several models and methodologies regarding the handling of evidence and digital forensic procedures during an investigation. These include the DFRWS model \cite{palmer2001road}, the Enhanced Digital Investigation Process Model
(EDIP) \cite{baryamureeba2004enhanced}, the Analytical Crime Scene Procedure Model (ACSPM) \cite{BULBUL2013244}, the integrated digital forensic process model of Kohn et al. \cite{KOHN2013103}, the Systematic digital forensic investigation model (SRDFIM). \cite{agarwal2011systematic}, and the advanced data acquisition model (ADAM) \cite{adams2013advanced}. In general, law enforcement agencies follow variants of the ACPO (Association of Chief Police
Officers) guidelines \cite{williams2012acpo}.

\begin{figure*}[!ht]
 \includegraphics[width=.75\textwidth]{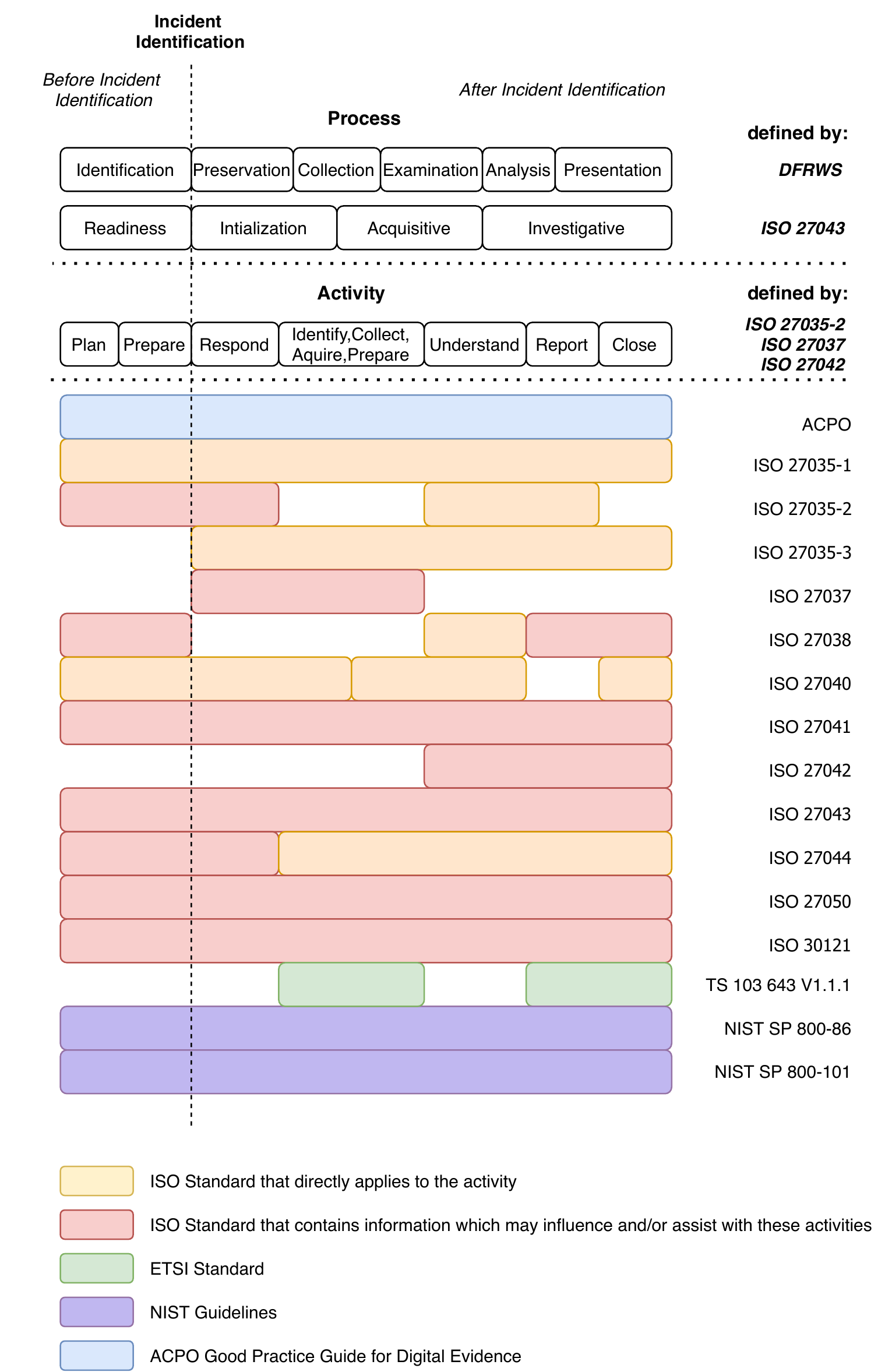}
 \centering
 \caption{Applicability of standards and guidelines to the investigation process classes and activities.}
 \label{fig:ISO-DF}
\end{figure*}

The International Organization for Standardization (ISO) has released a series of standards to assist in this effort by providing the family of ISO 27000, that focus on information security standardisation procedures. In what follows, we present the most relevant standards about digital forensics investigation of drones as described in Figure \ref{fig:ISO-DF} derived from ISO/IEC 27041:2015 \cite{iso_27041}.

\begin{itemize}

\item \textbf{ISO/IEC 27035:} This is a three-part standard that provides organisations with a structured and planned approach to the management of security incident management covering a range of incident response phases

\item \textbf{ISO/IEC 27037:2012:} This standard provides general guidelines about the handling of the evidence of the most common digital devices and \textit{the circumstances including devices that exist in various forms}, giving the example of an automotive system \cite{iso_27037} that uses many sensors that a drone does. This means that although the framework for the supporting devices of a drone exists in the ISO standard, the handling of the actual flying vehicle is not yet established.

\item \textbf{ISO/IEC 27038:2014:} Describes the digital redaction of information that must not be disclosed, taking extreme care to ensure that removed information is permanently unrecoverable.

\item \textbf{ISO/IEC 27040:2015:} Provides detailed technical guidance on how organisations can define an appropriate level of risk mitigation by employing a well-proven and consistent approach to the planning, design, documentation, and implementation of data storage security.

\item \textbf{ISO/IEC 27041:2015:} Describes other standards and documents to provide guidance, setting the fundamental principles ensuring that tools, techniques and methods, appropriately selected for the investigation.

\item \textbf{ISO/IEC 27042:2015:} This standard describes how methods and processes to be used during an investigation can be designed and implemented to allow correct evaluation of potential digital evidence, interpretation of digital evidence, and effective reporting of findings.

\item \textbf{ISO/IEC 27043:2015:} It defines the key common principles and processes underlying the investigation of incidents and provides a framework model for all stages of investigations.

\item \textbf{ISO/IEC 27050:} This recently revised standard guides non-technical and technical personnel to handle evidence on electronically stored information (ESI).

\item \textbf{ISO/IEC 17025:2017:} This standard describes the general requirements for the competence of testing and calibration laboratories, applying to all organisations performing laboratory activities, regardless of the number of personnel.

\item \textbf{ISO/IEC 30121:2015:} Provides a framework for organizations to strategically prepare for a digital investigation before an incident occurs, to maximise the effectiveness of the investigation.

\end{itemize}

ETSI is a European Standards Organization that produces standards for ICT systems and services that are used all around the world, collaborating with numerous organisations. In 2020 ETSI published TS 103 643 V1.1.1 (2020-01) \cite{etsi_2020}, a set of techniques for assurance of digital material in a legal proceeding, to provide a set of tools to assist the legitimate presentation of digital evidence
\footnote{\url{https://www.swgde.org/documents/published}}.

The National Institute of Standards and Technology (NIST) has released guidelines for organisations to \textit{develop forensic capability}, based on the principles of forensic science in the aspect of the application of science to the law but should not be used on digital forensic investigations due to subjection to different laws and regulations, as clearly stated on their manual. The scope on NIST guidelines is \textit{incorporating forensics into the information system life cycle} of an organisation. The most relevant guidelines are 800-86 \cite{10.5555/2206298} for Integrating Forensic Techniques into Incident Response and 800-101 \cite{ayers_brothers_jansen_2014} for Mobile Device Forensics.

The Scientific Working Group on Digital Evidence (SWGDE) is an organisation engaged in the field of digital and multimedia evidence to \textit{foster communication and cooperation as well as to ensure quality and consistency within the forensic community}. SWGDE has released a number of documents to provide the current best practices on a large variety of state of the art forensics subjects. Nonetheless, none of them is targeting nor addressing the particularities of drone forensics.

Nevertheless, in none of the aforementioned standards is there any reference to drones. While someone could oversee this omission as the procedures could cover the procedures needed for UAV forensic investigations. However, for the collection of forensic evidence from a UAV one cannot always access the data and may need to exploit the device to gain access to them. As will be discussed in the next paragraphs, such processes are not foreseen by any standard. Moreover, the contradicting records in the log files of the UAV and the GCS; also discussed in the following paragraphs, are not considered by any of the standards mentioned above. Therefore, while the equipment and the preparation of a forensic lab that would investigate a drone are foreseen, the particularities of drones are not covered by existing standards.

Finally, INTERPOL recently released an overview of drones and associated devices and guidance for digital forensic investigators responsible for the acquisition, examination, analysis, and presentation of the digital evidence from the drone \cite{inter}. This work tried to gap the knowledge of law enforcement officers on this state of the art forensics topic and is the first official organisation attempt towards a standardised procedure.

\subsection{Challenges and gaps in established procedures and current standard limitation}
Due to the nature of a drone, it is apparent that it is a system that significantly differentiates from traditional devices that an investigator would have to perform digital forensics. Firstly, it is a cyber-physical system; therefore, the physical part introduces many issues in the investigation. This can be manifested by various means, e.g. the sensors may have significant differences due to calibration or accuracy as in the case of the GPS sensor. To highlight this significant issue, the following should be considered. The accuracy of the GPS depends on the sensor's technology\footnote{\url{https://marxact.com/support/what-is-the-difference-between-standalone-differential-gps-and-real-time-kinematic/}}. A Standard GPS or has an accuracy of \textit{1.5 meters}, a Differential GPS \textit{40 centimeters} where in the case of RTK \textit{1 centimeter}. This deviation means that the GCS does not report the exact location of the UAV. In a case of a catastrophic event (e.g. collision with a nearby object or extreme weather conditions) the drone operator can \textit{only assume} the position of the drone, as reported by the GCS, and the data for the forensic investigation cannot be precisely evaluated. The accuracy of the sensor is not the only problem regarding satellite navigation systems. Drones are susceptible to GPS spoofing \cite{tippenhauer2011requirements,kerns2014unmanned,gaspar2018capture,arteaga2019analysis}, meaning that the validity and integrity of the GPS signal can be in question during an investigation.

As previously mentioned, the UAS consists of multiple devices, and it is not always clear which one would be available during a forensic investigation. For instance, the UAV may have been collected by the incident responder but may not have access to the GCS device either because it was destroyed or not found during the investigation. The inverse may also apply, that is the UAV cannot be recovered, and the investigator has access only to the GCS device. Therefore, there is no full access to possible evidence sources. Despite the lack of a source of evidence, a big challenge is to determine which source should be trusted in the case that the two sources have contradictory or missing logs. Note that the differences may stem from various reasons as the problem is amplified by the lack of proper authentication mechanisms. Since in most UAVs the commands are not authenticated via cryptographic primitives one may craft the log files in any of the two devices. Even if this is not the case, the differences from the sensors (see the discussion above for the differences in GPS) introduce further issues for the investigator.

Currently, there are also numerous drone models that lack essential security features that may hinder the forensic investigation, into formally presenting the case. The lack of encryption and authentication on commercially available products is a profound example. Due to their absence, the received commands are not authenticated, and their source can be questioned. Therefore, it is not evident who issued a command to the drone, that in the case of an autonomous mission could significantly deviate from the expected flight path or pattern. On the contrary, when there is encryption used on the device, the key may not always be available, so the acquisition from forensics tools might not be possible.

The latter issue with encryption leads to another challenge. To collect the evidence from a drone that lacks accessible removable storage devices, or they are encrypted, similar to mobile devices, one may need to exploit a vulnerability to gain access to the data. This implies tampering of the digital evidence introducing issues for the chain of custody of the evidence. The nature of the exploit and the need to, e.g. ``root'' the device may significantly alter the stored information and careful examination of the tools and methods is necessary to allow the presentation of the findings in a court of law. Clearly, such practices violate the first principle of ACPO, which clearly states that:
\begin{quote}
 ``\textit{No action taken by law enforcement agencies or their agents should change data held on a computer or storage media which may subsequently be relied upon in court.}''
\end{quote}

One has also to consider the diversity of the platforms. Unlike mobile phones where there are two main platforms, Android and iOS, and can be considered a well-established market the drone market might have a big player (DJI) which enjoys the biggest market share; more than 70\%, nevertheless, the rest is significantly fragmented. Therefore, there are plenty of different platforms which imply the need for various forensics tools. Furthermore, the different scope of many of the drones as well as their big numbers suggest that despite the fragmentation this heterogeneity is expected to continue, so there is no foreseen interoperability of the forensics tools. It is worth noticing that the recent ban on Chinese manufactured drones may lead to significant market changes\footnote{\url{https://www.ainonline.com/aviation-news/aerospace/2020-08-17/us-nears-law-banning-chinese-drones}}.

One very challenging parameter for drone forensics has to do with live forensics. While
Adelstein \cite{adelstein2006live} stressed the need for live forensics, we have to acknowledge the additional difficulties and limitations of performing live digital forensics in a drone. Contrary to ``traditionally'' compromised devices where the adversary will have integrated evasion, persistence and anti-analysis mechanisms, in a drone the adversary may have installed additional physical mechanisms, so the drone may physically react, e.g. consider it being armoured, or try to fly away. Beyond that, since the drone is remotely controlled; even via the cloud, kill switches or other such mechanisms are expected to continue being relevant during the investigation. To this end, to capture a drone, it is advised to approach it from a non-visible angle and use a non-tampering measure such as a coat or net. Next, the drone should be turned off without tampering any possible wet evidence on it.

Since there is no standardised procedure to conduct a digital forensics investigation on drones, researchers have to resolve to already established practices that apply to similar devices. This raises the following question: What kind of device is a drone or relevant to, to select a method to proceed with the forensics triage. A valid argument would be that a drone has many similarities with a mobile phone regarding the hardware (e.g. processor architecture) and the sensors (e.g. GPS receiver, accelerometer). It should be noted here that there is also no standard for handling mobile digital forensics case, yet an initiative has recently started by the FORMOBILE H2020 project\footnote{\url{https://formobile-project.eu/}}.

\begin{figure*}[th]
 \includegraphics[width=.8\textwidth]{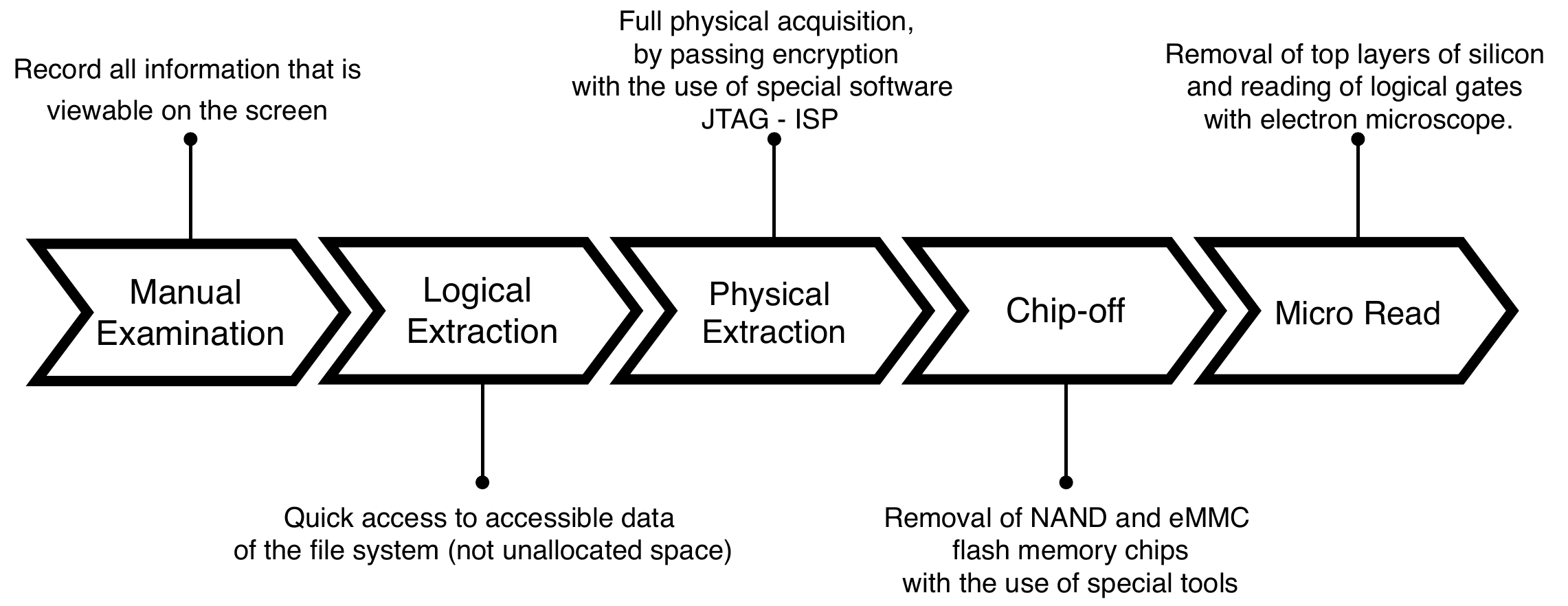}
 \centering
 \caption{Data acquisition techniques.}
 \label{fig:ac-methods}
\end{figure*}

To acquire the needed data, the investigator will have to resolve to the techniques illustrated in Figure \ref{fig:ac-methods}. Typically to \textit{logically image} a mobile device, the investigator has to turn on the device. If this method is applied to a drone, there is a high risk to alter or erase the existing data of the UAV's internal memory \cite{arshad2018digital}. As stated in \cite{DROP}, turning on the drone will erase the previously recorded flight log, resulting in the loss of evidence and alteration of the investigation. Although the investigator followed the procedure correctly, the result may render the whole investigation useless due to misconceptions about the drone's internal operations. Since the storage of data on the drone itself is highly volatile, extreme caution and preparation for the acquisition should take place. Therefore, the investigator should take into consideration which is the device from which she has to collect evidence and under which mode it allows their extraction to prevent tampering with them.

This example raises another concern about the technical knowledge of the investigators tasked with acquiring data from the drone. Although the investigators are acquainted with procedures to handle different kind of devices, the cognitive and human factors \cite{sunde2019cognitive} coupled with the lack of expertise on the drones' technical details, that comes on the standardisation of the process, is of great importance. When deemed necessary to switch on the device, it is advised that the corresponding environment has Faraday shielding to prevent network signals from or to the drone. If this is not possible, then jamming devices or aluminium foils must be used to prevent incoming and outgoing network traffic from the drone.

Moreover, there are several legal and ethical aspects that underpin such a forensic investigation and may introduce several restrictions for the investigator. For instance, depending on the jurisdiction, the investigator may not be allowed to use the stored credentials in the drone to collect further evidence.

\section{Conclusions}
Due to the continuous adoption of UAVs in various domains from shipping and delivery to agriculture, and from disaster management and smart city monitoring to search and rescue operations, the skies are being populated continuously with more and more flying vehicles. The fact that UAVs can save both time and costs while offering novel services implies that this trend will continue. Moreover, we have already witnessed their malicious usage in various acts, both physical and digital. Based on the above one can easily understand that the increase in the numbers of drones coupled with their exploitation from adversaries is going to soon make digital forensics investigations on drones a norm. Insurance companies, law enforcement agencies, security firms and individuals will have to extract evidence from a drone to investigate a case. Nonetheless, as already discussed in this work, a UAV is very different compared to traditional computing devices. In fact, its physical nature, its mobility, and duality; in terms of control, create a very complex environment. The latter aspects are not covered by existing standards, further perplexing the investigations.

We argue that it is a high time for the corresponding stakeholders to acknowledge these issues and collaborate to develop some baseline principles that will facilitate the investigations in the near future. The challenges are multiple and cannot be solved individually with many of them requiring further steps from the manufacturers. A typical example can be considered the use of cryptographic primitives to authenticate the received commands from the GCS. It is straightforward that this measure increases the overall security of a drone, preventing many command injection attacks; nevertheless, it provides the guarantees of the audit trail for the forensic investigator. Further unification of interfaces from the manufacturers may also facilitate the development of forensic tools that address the needs for more platforms.

Several commonalities of UAVs and mobile phones in terms of extraction of evidence require a more holistic approach from the forensics investigators for both cases. As discussed, one may need to exploit some vulnerability to gain access to such devices to gain access to the evidence which violates the principles of ACPO and many existing standards. This highlights a critical issue in the collection of digital evidence as it questions the chain of custody of the evidence. Evidently, it opens the door for exploitation for multiple perspectives if it is not correctly addressed. For instance, one may use it to render the evidence useless or one may use it to implant evidence. Evidently, the latter indicates that a legal perspective is deemed necessary. Notably, we have also identified jurisdiction issues that require the legal approach.

Based on our gap analysis, we believe that the development of a standard for drone forensics is necessary and timely. The aforementioned challenges require several fundamental changes in the drones per se and the established forensics procedures and more coordinated efforts towards this direction. To this end, we prioritise the harmonisation of manufacturers with common best practices from other platforms that will pave the way for procedures and tools could be established and developed.

\section*{Acknowledgements}
This work was supported by the European Commission under the Horizon 2020 Programme (H2020), as part of the projects \textit{CyberSec4Europe} (Grant Agreement no. 830929) and \textit{LOCARD} (Grant Agreement no. 832735).

The content of this article does not reflect the official opinion of the European Union. Responsibility for the information and views expressed therein lies entirely with the authors.
\bibliographystyle{plain}
\bibliography{refs}
\end{document}